\documentclass[aps,superscriptaddress,showpacs]{revtex4-2}
\usepackage{amssymb,bm}
\usepackage{graphicx}
\usepackage{amsmath}
\usepackage{epstopdf}
\usepackage{comment}
\allowdisplaybreaks
\graphicspath{{Figs/}}

\begin{document}
	
	\title{Non-trivial solution to a simple problem	}
		\author{A.I.Milstein}\email{A.I.Milstein@inp.nsk.su}

\affiliation{\textit{Budker Institute of Nuclear Physics, 630090, Novosibirsk,Russia}}
\affiliation{\textit{Novosibirsk State University, 630090, Novosibirsk, Russia}}

\begin{abstract}
	The problem of finding the frequencies of  small longitudinal oscillations of a spring having a finite mass and stiffness, attached at one end to a  wall and at the other end to a body of finite mass, is discussed. This problem was repeatedly proposed at Olympiads for schoolchildren, in various lessons on the Internet, and even  on tests in mechanics for students of universities. In all the cases known to me, the implied solution was actually wrong. I discuss two cases: (A) a spring lies on a smooth table, (B) a spring is attached to the ceiling. It is shown that the solution to this simply formulated problem is non-trivial.
	\end{abstract}

\maketitle

\section{Introduction}
\begin{figure}
	\centering
	\includegraphics[totalheight=6cm]{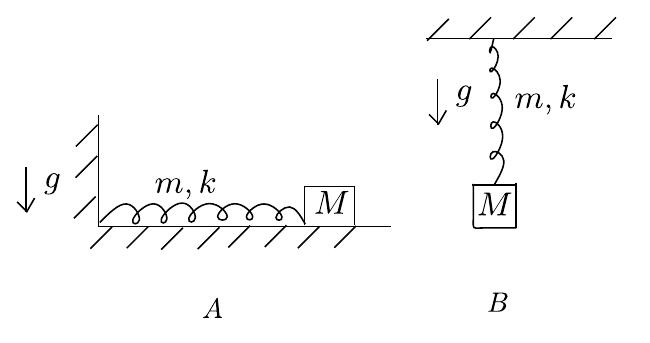}
	\caption{Oscillations of a spring having a finite mass $m$ and stiffness $k$, $g$ is the acceleration of gravity. (A) The spring is on a horizontal smooth plane. (B) The spring is attached to the ceiling.}\label{figab}
\end{figure}

The reason to write this article was a problem that I accidentally came across on the Internet. It turned out that this problem was very popular and was discussed many times in various online lessons, offered at school Olympiads and even on tests in mechanics for students of universities. To my surprise, a solution implied to be  correct was actually wrong. The problem is formulated very simply (see Fig.\ref{figab}A): find the frequency of small longitudinal oscillations of a spring having a finite mass $m$ and stiffness $k$, attached at one end to a  wall and at the other end to a body of mass $M$; the spring is on a horizontal plane,  friction should be neglected.

The solution that was considered to be correct was the following. Let $u$ is the speed of the body and the length of undeformed spring is $l$. Then, the speed of a point located at a distance $y$ from the fixed end is  $(y/l)u$. Accordingly, the total kinetic energy of the system is 
$$ T= \dfrac{1}{2}Mu^2+\dfrac{1}{2} \int_0^l \left(\dfrac{m}{l}\right)\cdot \dfrac{y^2}{l^2}u^2\,dy=
\dfrac{1}{2}\left(M+m/3\right)\,u^2\,.$$
The potential energy $U$ of a system is  $U=kz^2/2$, where $z$ is the deviation of the body from the equilibrium  position, $u=\dot z\,$. Comparing $T$ and $U$ with the corresponding expressions for a conventional spring oscillator, we find the oscillation frequency $\Omega$:
\begin{equation}\label{Omega}
	\Omega=\sqrt{\dfrac{k}{M+m/3}}\,.
\end{equation}

The above solution cannot be correct, because there should be an infinite number of oscillation frequencies, but not one. The set of functions corresponding to different modes of oscillation should form a complete set of functions which allows one to find a solution for a spring motion   with an attached body under arbitrary initial conditions. This paper is devoted to finding these oscillation modes. I  show that Eq.~\eqref{Omega} is not completely meaningless, since it represents a good approximation to one of  frequencies for an arbitrary value of the ratio $Q=m/M$. The case of a spring attached to the ceiling is also considered (see Fig.\ref{figab}B).

\section{Equation of oscillations }\label{seceq}
Let a point with coordinate $y$ in an undeformed spring of length $l$ deviate from its position by a distance $x(y,t)$ at time $t$.
As is known, one-dimensional oscillations of an elastic medium are described by the wave equation
\begin{equation}\label{weq}
	\ddot x(y,t)=c^2\, x''(y,t)\,,
\end{equation}	
where $\ddot x(y,t)=(\partial^2/\partial t^2)\,x(y,t)$, 	$x''(y,t)=(\partial^2/\partial y^2)\,x(y,t)$  and $c$  is the speed of sound. In our case
\begin{equation}\label{ceq}
	c=\sqrt{\dfrac{k}{m}}\,l\,.
\end{equation}	
It is not difficult to derive Eqs.~\eqref{weq} and \eqref{ceq} if we represent a spring with mass $m$ and stiffness $k$ as $N$ balls with masses $m_0=m/N$, interconnected by springs having stiffness $k_0=kN$ and length $l_0=l/N$, with the first ball connected by a spring to the wall, and the last ball to a body of mass $M$. It is assumed that $N\gg 1$. Let the deviation of the ball with number $n$ at time $t$ from its equilibrium position be equal to $x_n(t)$. Write down Newton's second law for the motion of this ball:
\begin{equation}\label{neq}
	m_0\,\ddot x_n(t)= k_0 [(x_{n+1}-x_n)+ (x_{n-1}-x_n)] \,.
\end{equation} 
Since $x_{n+1}(t)-x_n(t)$ is a small quantity for $N\gg 1$, then
\begin{align}\label{difeq}
	&x_{n+1}(t)-x_n(t)\approx \dfrac{\partial}{\partial n}\,x_n(t)+\dfrac{1}{2}\,\dfrac{\partial^2}{\partial n^2}\,x_n(t)= l_0\dfrac{\partial}{\partial y}x(y,t)+\dfrac{l_0^2}{2}\,\dfrac{\partial^2}{\partial y^2}x(y,t)\,,\nonumber\\
	&x_{n-1}(t)-x_n(t)\approx -\dfrac{\partial}{\partial n}\,x_n(t)+\dfrac{1}{2}\,\dfrac{\partial^2}{\partial n^2}\,x_n(t)= -l_0\dfrac{\partial}{\partial y}x(y,t)+\dfrac{l_0^2}{2}\,\dfrac{\partial^2}{\partial y^2}x(y,t)\,,
\end{align}
where $y=n l_0$ and $x_n(t)=x(y,t)$. Substituting \eqref{difeq} into \eqref{neq}, we obtain the equation
$$
\ddot x(y,t)=\dfrac{k_0 }{m_0}l_0^2\, x''(y,t)=\dfrac{k }{m}l^2\, x''(y,t)\,, $$
coinciding with \eqref{weq}. Next, let the deviation of a body of mass $M$ from the equilibrium position is $x_{N+1}(t)$. Then Newton's second law for body motion has the form
$$M\,\ddot x_{N+1}(t)= k_0 \, (x_{N}-x_{N+1})=-k_0\,\dfrac{\partial}{\partial n}\,x_{n}(t)|_{n=N+1}=-k_0l_0\,\dfrac{\partial}{\partial y}\,x(y,t)|_{y=l}\,.$$
Taking into account the relation $k_0l_0=kl$, we find the boundary condition
\begin{equation}\label{graneq}
	M\,\ddot x(l,t)= -kl\,\dfrac{\partial}{\partial y}\,x(y,t)|_{y=l}  \,.
\end{equation}
Thus, to find the frequencies of small oscillations, one needs to solve Eq.~\eqref{weq} with boundary conditions \eqref{graneq} and
\begin{equation}\label{gran0eq}
	x(0,t)=0\,.
\end{equation}
The derivation of equations given here, based on discretization, is similar to that used in many textbooks when discussing sound oscillations (see, e.g., the textbook \cite{ziman}).

\section{Calculation of frequencies}
\begin{figure}
	\centering
	\includegraphics[totalheight=6cm]{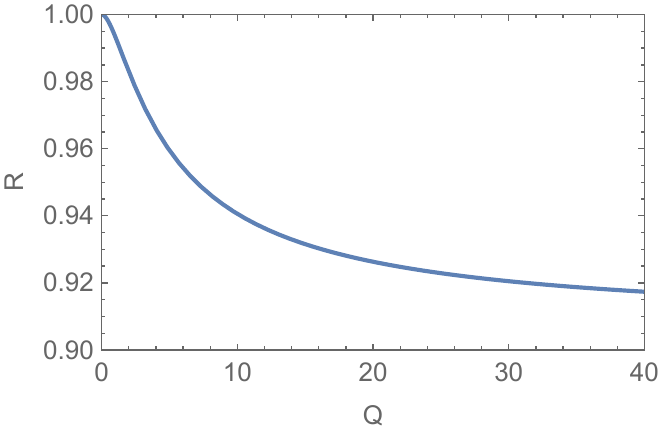}
	\caption{Dependence of $R=\omega_0/\Omega$ on $Q=m/M$.}\label{R}
\end{figure}

It follows from the boundary condition \eqref{gran0eq}  that the solution to the wave equation with a certain frequency $\omega_j$ is standing wave:
\begin{equation}\label{xi}
	\eta_j(y,t)=A_j\sin(\omega_j t+\alpha_j)\,\sin(q_j y)  \,,\quad q_j=\dfrac{\omega_j}{c}\,,
\end{equation}
where $A_j$ and  $\alpha_j$ are the amplitude and phase of the wave. From the boundary condition~ \eqref{graneq} we find the dispersion equation
\begin{equation}\label{phi}
	\varphi_j\,\mbox{tg}\varphi_j=Q\,,
\end{equation}
where $Q={m}/{M}$ and $\varphi_j={\omega_j\,l}/{c}$.
Using \eqref{ceq}, we obtain an expression for the frequency
\begin{equation}\label{omegai}
	\omega_j=\varphi_j\,\sqrt{\dfrac{k}{m}}\,.
\end{equation}
Eq.~\eqref{phi} depends only on the ratio $Q$ and is independent of $k$ and $l$. As it should be,  Eq.~\eqref{phi} has an infinite number of solutions $\varphi_j$, to which the frequencies $\omega_j$ are proportional. These solutions belong to the intervals
$$j\pi< \varphi_j< j\pi +\pi/2\,,$$
where $j=0,1,\dots$.  If $Q\ll 1$, then $\varphi_0=\sqrt{Q}$ and $\varphi_{j>0}=j\pi+Q/(j\pi)$. Therefore, for $Q\ll 1$
\begin{equation}
	\omega_0=\sqrt{\dfrac{k}{M}}\,,\quad \omega_{j>0}=j\pi\,\sqrt{\dfrac{k}{m}}\,\gg \omega_0\,.
\end{equation} 
If $Q\gg 1$, then $\varphi_j=(j\pi+\pi/2)(1-1/Q)$. As a result, for $Q\gg 1$
\begin{equation}\omega_j=(j\pi+\pi/2)\,\sqrt{\dfrac{k}{m}}\,.
\end{equation}
It is natural to compare the frequency $\omega_0$ with $\Omega$ \eqref{Omega}. We see that
$\omega_0/\Omega=1$ for $Q\ll 1$ and $\omega_0/\Omega=\pi/(2\sqrt{3})=0.907$ for $Q\gg 1$, so that 
$\omega_0$ and $\Omega$ are practically the same for all values of $Q\,$! The dependence of  ratio $R=\omega_0/\Omega$ on $Q$ is shown in Fig.~\ref{R}.
\section{Problem with initial conditions.}
As already noted, an infinite number of frequencies is necessary to solve the problem with initial conditions 
$$x(y,0)=X(y)\,,\quad \dot x(y,0)=V(y)\,.$$
Let us write the corresponding solution $x(y,t)$ in the form
\begin{equation}\label{sol}
	x(y,t)=\sum_{j=0}^\infty\, [a_j\cos(\omega_j t)+b_j\sin(\omega_j t)]\,\sin(q_j y)  \,,\quad q_j=\dfrac{\omega_j}{c}\,.
\end{equation} 
It follows from the boundary conditions  that
\begin{align}\label{gran1}
	&	X(y)=\sum_{j=0}^\infty\, a_j\,\sin(q_j y) \,,\quad
	V(y)=\sum_{j=0}^\infty\,\omega_j\, b_j\,\sin(q_j y) \,.
\end{align} 
To find the coefficients $a_j$ and $b_j$ from these equations, we use the relations
\begin{align}\label{rel}
	&\int_0^l\cos(q_i y) \cos(q_j y)\,dy=0\,,\quad i\neq j\,;\nonumber\\
	&\int_0^l\cos^2(q_j y)\,dy=\dfrac{l}{2}\left(1+\dfrac{\sin2\varphi_j}{2\varphi_j}\right) \,.
\end{align} 
Taking the derivative with respect to $y$ from both sides of Eqs.~\eqref{gran1} and applying the relations \eqref{rel}, we obtain
\begin{align}
	&a_j=\dfrac{4}{2\varphi_j+\sin2\varphi_j}\int_0^l\, \cos(q_j y)\,X'(y)\,dy\,, \nonumber\\
	&\omega_j\,b_j=\dfrac{4}{2\varphi_j+\sin2\varphi_j}\int_0^l\, \cos(q_j y)\,V'(y)\,dy\,.
\end{align} 
Interestingly, solutions with different $j$ are not orthogonal to each other in the common sense:
\begin{align}
	&\int_0^l\sin(q_i y) \sin(q_j y)\,dy=-\dfrac{l}{Q}\,\sin\varphi_i\, \sin\varphi_j\neq0\,,\quad i\neq j\,.
\end{align} 

Let us consider an example of the boundary conditions $X(y)=Ly/l$ and $V(y)=0$, corresponding to the experiment in which we slowly stretched the spring to a length $L$ and then released it. In this case $b_j=0$ and
\begin{align}
	&a_j=\dfrac{4 \sin\varphi_j}{(2\varphi_j+\sin2\varphi_j)\,\varphi_j}\,L\,.
\end{align} 
At $Q\ll 1$ we have
\begin{equation}
	a_0=\dfrac{L}{\sqrt{Q}}\,,\quad a_{j>0}=\dfrac{2\,(-1)^j\,Q}{(j\pi)^3}\,L\,\ll a_0\,.
\end{equation} 
Thus, at $Q\ll 1$ higher modes are not excited. For $Q\gg 1$ we find
\begin{equation}
	a_{j}=\dfrac{2\,(-1)^j}{(j\pi+\pi/2)^2}\,L\,.
\end{equation} 
Consequently, at $Q\gg 1$ all modes are excited, but the wave amplitudes quickly decrease with increasing $j\,$.

It is interesting to discuss the fraction of the total energy $E$ in each mode. Using the discrete model described in Sec.~\ref{seceq}, we obtain
\begin{equation}
	E=\dfrac{m}{2l}\,\int_0^l\,\dot x^2(y,t)\,dy+\dfrac{kl}{2}\,\int_0^l\, x'^2(y,t)\,dy+\dfrac{M}{2}\dot x^2(l,t)\,.
\end{equation} 
Substituting Eq.~\eqref{sol} into this formula, we find
\begin{align}
	&E=\sum_{j=0}^\infty\, \varepsilon_j \,,\quad \varepsilon_j =\dfrac{k}{8}\,\varphi_j\left(2\varphi_j+\sin2\varphi_j\right)\,(a_j^2+b_j^2)\,.
\end{align} 
For the case considered above, $E=kL^2/2$ and
\begin{align}
	& \varepsilon_j =\dfrac{kL^2}{2}\,\dfrac{4\sin^2\varphi_j}{\varphi_j\left(2\varphi_j+\sin2\varphi_j\right)}\,.
\end{align} 
For $Q\ll1$ we have
$$\dfrac{(E-\varepsilon_0)}{E}=\dfrac{Q^2}{45}\ll1\,.$$
For $Q\gg1$ the contribution of oscillations with $\omega_{j>0}$ to the energy  is 
$$\dfrac{(E-\varepsilon_0)}{E}=1-\dfrac{8}{\pi^2}=0.19\,.$$
Thus, in the example under consideration, the main contribution to the energy comes from the oscillation mode with the lowest frequency.
It is easy, however, to come up with a problem where the oscillation mode with an arbitrary frequency $\omega_j$ is mainly excited. To do this, forced oscillations must be used.

\section{Forced oscillations}
Let us apply a periodic force ${\cal F}(t)=F\sin(\nu t)$ to the body. Then we have the equations
\begin{align}\label{eqF}
	\ddot x(y,t)=c^2\, x''(y,t)\,,\quad M\,\ddot x(l,t)= -kl\,\dfrac{\partial}{\partial y}\,x(y,t)|_{y=l} +
	F\sin(\nu t)\,.
\end{align}	
For $\nu\neq \omega_j$ the solution corresponding to forced oscillations is 
\begin{align}\label{solF}
	x(y,t)=A\sin(\nu t)\,\sin\left(\dfrac{\nu}{c}y\right)\,,\quad 
	A=\dfrac{F}{k}\,\dfrac{Q}{\phi\cos\phi\,(Q-\phi\,\mbox{tg}\phi)}\,,
\end{align}	
where $\phi=\nu l/c$. It is seen that at $\phi\rightarrow \varphi_j$ the amplitude $A$ of forced oscillations  tends to infinity (resonance). For $\nu= \omega_j$ the resonance solution reads
\begin{align}\label{solR}
	x(y,t)=-\dfrac{2F}{kl}\,\dfrac{\sin\varphi_i}{2\varphi_i+\sin2\varphi_i}\,[ct\,\cos(\omega_it)\sin(q_iy)+y\,\sin(\omega_it)\cos(q_iy)]\,,
\end{align}	
where $q_i=\omega_i/c$. As it should be, the amplitude of  oscillations increases linearly with time.

\section{Spring is attached to the ceiling.}

Let us now find the oscillation frequencies of a spring attached to the ceiling (see Fig. \ref{figab}B). Similar to the derivation given in Sec.~\ref{seceq}, we obtain the equations
\begin{align}\label{eqh}
	\ddot x(y,t)=c^2\, x''(y,t)+g\,,\quad 	M\,\ddot x(l,t)= -kl\,\dfrac{\partial}{\partial y}\,x(y,t)|_{y=l} +Mg \,,
\end{align}	
where $g$ is the acceleration of gravity and $c^2=kl^2/m$. The stationary (time independent) solution $h(y)$ of these equations  has the form
\begin{align}\label{eqX}
	h(y)=\dfrac{g}{k}\left[(m+M)\,\dfrac{y}{l}-m\,\dfrac{y^2}{2\,l^2}\right] \,.
\end{align}	
Making substitution $x(y,t)=h(y)+ \widetilde x(y,t)$ we find that the equation for the function
$\widetilde x(y,t)$ coincides with Eq.~\eqref{eqh} at $g=0$. Thus, the frequencies of small oscillations in cases (A) and (B) are the same.

\section{Conclusion}
This paper shows that the seemingly simple problem of oscillations of a massive spring with a body of finite mass attached to it has a non-trivial solution that has interesting physical content. Using the example considered, we are once again convinced that not everything that seems obvious is correct.


\begin{thebibliography}{99}
	\bibitem{ziman}
	
J.M. Ziman, \textit{Principles of the theory of solids} (Cambridge, The University Press, 1974). 
	
\end{thebibliography}
\end{document}